\documentclass[prd, twocolumn, nofootinbib, tightenlines,preprintnumbers,notitlepage,longbibliography,superscriptaddress]{revtex4-1}
\usepackage{graphicx}
\usepackage{amsmath}
\usepackage{amsfonts}
\usepackage[colorlinks=true,citecolor=blue,urlcolor=cyan,linkcolor=black]{hyperref}
\usepackage{tablefootnote}
\usepackage{multirow}

\usepackage{aas_macros}

\usepackage{setspace}
\usepackage[hang]{footmisc} 
\usepackage{scalerel,stackengine}
\stackMath
\newcommand\what[1]{%
\savestack{\tmpbox}{\stretchto{%
  \scaleto{%
    \scalerel*[\widthof{\ensuremath{#1}}]{\kern-.6pt\bigwedge\kern-.6pt}%
    {\rule[-\textheight/2]{1ex}{\textheight}}%WIDTH-LIMITED BIG WEDGE
  }{\textheight}% 
}{0.5ex}}%
\stackon[1pt]{#1}{\tmpbox}%
}
\usepackage{tikz} 
\usetikzlibrary{shapes,arrows,positioning,automata,backgrounds,calc,er,patterns}
\usepackage[compat=1.1.0]{tikz-feynman}
\usepackage{graphicx}
\usepackage{dcolumn}
\usepackage{bm}
\usepackage[normalem]{ulem}
\usepackage{xcolor}

\graphicspath{ {figures/} }

\usepackage{epstopdf}
\newcommand{\be}{\begin{eqnarray}}
\newcommand{\non}{\nonumber \\}
\newcommand{\ee}{\end{eqnarray}}

\newcommand{\bk}{\boldsymbol{k}}

\usepackage[export]{adjustbox}% http://ctan.org/pkg/adjustbox

\def\p{\parallel}
\def\hr{{\hat r}}
\def\hk{{\hat k}}
\def\hv{{\hat v}}
\def\k{\boldsymbol{k}}

\newcommand{\vpot}{\Upsilon}

\def\rhat{\hat{\boldsymbol{r}}}

\usepackage{suffix}
\usepackage{mathtools}

\DeclarePairedDelimiterX\MeijerM[3]{\lparen}{\rparen}%
{\begin{smallmatrix}#1 \\ #2\end{smallmatrix}\delimsize\vert\,#3}

\newcommand\MeijerG[8][]{%
  G^{\,#2,#3}_{#4,#5}\MeijerM[#1]{#6}{#7}{#8}}

\WithSuffix\newcommand\MeijerG*[7]{%
  G^{\,#1,#2}_{#3,#4}\MeijerM*{#5}{#6}{#7}}

\newcommand{\dd}{{\rm d}}

\begin{document}

\title{Cosmology with the moving lens effect}

\newcommand{\imperial}{Department of Physics, Imperial College London, Blackett Laboratory, Prince Consort Road, London SW7 2AZ, UK}

\newcommand{\jhu}{Department of Physics \& Astronomy, Johns Hopkins University, Baltimore, MD 21218, USA}

\newcommand{\perimeter}{Perimeter Institute for Theoretical Physics, 31 Caroline St N, Waterloo, ON N2L 2Y5, Canada}

\author{Selim~C.~Hotinli}
\affiliation{\jhu}

\author{Kendrick~M.~Smith}
\affiliation{\perimeter}

\author{Mathew S.~Madhavacheril}
\affiliation{\perimeter}

\author{Marc~Kamionkowski}
\affiliation{\jhu}

\date{\today}

\begin{abstract}

Velocity fields can be reconstructed at cosmological scales from their influence on the correlation between the cosmic microwave background and large-scale structure. Effects that induce such correlations include the kinetic Sunyaev Zel'dovich (kSZ) effect and the moving-lens effect, both of which will be measured to high precision with upcoming cosmology experiments. Galaxy measurements also provide a window into measuring velocities from the effect of redshift-space distortions (RSDs). The information that can be accessed from the kSZ or RSDs, however, is limited by astrophysical uncertainties and systematic effects, which may significantly reduce our ability to constrain cosmological parameters such as $f\sigma_8$. In this paper, we show how the large-scale transverse-velocity field, which can be reconstructed from measurements of the moving-lens effect, can be used to measure $f\sigma_8$ to high precision.

\end{abstract}

\maketitle

\section{Introduction}

Next generation cosmic microwave background (CMB) experiments such as the Simons Observatory (SO)~\citep{Ade:2018sbj,Abitbol:2019nhf} and CMB-S4~\citep{Abazajian:2016yjj}, and galaxy surveys such as DESI~\citep{Aghamousa:2016zmz} and the Vera Rubin Observatory (VRO)~\citep{2009arXiv0912.0201L} will generate a wealth of new data with unprecedented precision on small scales. Correlations between CMB anisotropies and the galaxy density carry valuable cosmological information about the largest scales, creating novel opportunities for inference~\citep[e.g.][]{Seljak:2008xr,Zhang:2015uta,Banerjee:2016suz,Schmittfull:2017ffw,Modi:2017wds,Deutsch:2017ybc,Munchmeyer:2018eey,Cayuso:2019hen,Pan:2019dax,Hotinli:2019wdp}. These correlations that are induced by interactions of CMB photons with the intervening large-scale structure include the thermal and kinetic Sunyaev Zel'dovich effects~\citep{1969Ap&SS...4..301Z,1970A&A.....5...84Z,1980ARA&A..18..537S,1972CoASP...4..173S,Sazonov:1999zp} and the integrated Sachs-Wolfe effects~\citep{1967ApJ...147...73S}, which includes the moving-lens effect~\citep{1983Natur.302..315B}. Among these, the kinetic Sunyaev Zel'dovich (kSZ) and the moving-lens effects depend on the peculiar velocities of cosmological structures. 

A key product of velocity measurements is the combination of the linear-theory growth rate $f$ and the amplitude $\sigma_8$ of matter fluctuations on the scale of $8h^{-1}$Mpc. In principle, $f\sigma_8$ can be determined by measurements of the kSZ effect~\citep[e.g.][]{Deutsch:2017cja,Deutsch:2017ybc,Deutsch:2018umo,Smith:2018bpn,Munchmeyer:2018eey,Madhavacheril:2019buy,Contreras:2019bxy,Cayuso:2019hen,Hotinli:2020csk,Giri:2020pkk}, or by measurements of redshift-space distortions (RSDs) on galaxy surveys~\citep{Kaiser:1987qv}. The cosmological information available from the kSZ effect, however, is limited by the degeneracy of this signal with the optical depth of galaxies {(except in the case of primordial non-Gaussianity \cite{Munchmeyer:2018eey})}. While fast radio burst (FRB) sources may provide a way to break some of these biases by measuring the galaxy-electron correlation~\citep{Madhavacheril:2019buy}, the prospect of localising enough FRBs in the near future is challenging. 

Furthermore, recent studies show evidence of anisotropic selection effects~\citep[e.g.][]{Obuljen:2020ypy} for galaxy RSD measurements. Such effects are degenerate with redshift-space distortions, in the sense that the large-scale bias of the galaxy field becomes $(b_g\!+\!b_{\rm rsd}~\!f\mu^2)$, rather than simply $(b_g\!+\!f\mu^2)$, where $\mu\!=\!\hat{\boldsymbol{r}}\cdot\hat{\boldsymbol{k}}$. Here, $\rhat$ is the unit vector in the radial direction, $\hat{\bk}$ is the three-dimensional Fourier (unit) wavevector and $b_{\rm rsd}$ is a new bias parameter, defined such that $b_{\rm rsd}\!=\!1$ in the absence of anisotropic selection effects.\footnote{According to Ref.~\citep{Obuljen:2020ypy}, $(b_{\rm rsd}-1)$ can be of order $0.1-0.2$ in subcatalogs of SDSS.} Anisotropic selection effects could significantly degrade measurements of $f\,\sigma_8$ from future galaxy surveys, due to degeneracy with $b_{\rm rsd}$. In this paper, we propose a complementary method for measuring $f\,\sigma_8$ in which all degeneracies with astrophysical bias parameters are broken.

We will show that the temperature fluctuations sourced by the transverse velocities of large-scale structure (the moving-lens effect~\citep{1983Natur.302..315B}) can be used to measure $f\sigma_8$ and to break the kSZ optical-depth degeneracy. These transverse velocities are inferred by cross-correlation of the effect they induce in the CMB with galaxy surveys~\citep[e.g.][]{Hotinli:2018yyc,Yasini:2018rrl,Hotinli:2020ntd}. The moving lens effect sourced by individual objects is expected to be small and can be easily confused with other effects. However, combining the signal from many objects allows a tomographic reconstruction of the transverse-velocity field on large scales, analogous and complementary to kSZ tomography, which provides the radial-velocity fields~\citep[e.g.][]{Deutsch:2017cja,Smith:2018bpn}. Note that the transverse-velocities measured from moving-lens tomography will be biased similar to the case of kSZ tomography. The moving-lens bias $b_\perp$ can be measured through galaxy-galaxy or galaxy-CMB lensing, however, since $b_\perp$ is proportional to the galaxy-matter cross-correlation, in contrast to the kSZ bias $b_\p$ which is proportional to the galaxy-electron cross-correlation, which is difficult to measure independently without many localised FRB sources~\citep{Madhavacheril:2019buy}.

%similar to the standard kSZ tomography (i.e. radial-velocity reconstruction~\citep[e.g.][]{Deutsch:2017cja,Smith:2018bpn}). 

Small-scale CMB anisotropies are expected to be dominated by the kSZ effect due to Compton scattering of the CMB photon off free electrons that have non-zero peculiar velocities with respect to the CMB rest frame. This induces a shift in the CMB temperature in the direction of the free-electron radial velocity. The moving-lens effect is due to time-evolving gravitational potentials on the line of sight. This induces additional temperature anisotropies in the CMB, via the non-linear ISW (or Rees-Sciama) effect. This signal, which is a dipole pattern oriented along the object's transverse velocity, is expected to be smaller than kSZ. Even so, it will be measured with high signal-to-noise with upcoming CMB experiments~\citep{Hotinli:2018yyc,Yasini:2018rrl,Hotinli:2020ntd}. 

In what follows, we demonstrate that the transverse velocities from the moving-lens effect, when combined with galaxy measurements, can provide competitive constraints on the $f\sigma_8$ parameter, comparable to the scenario where the RSD bias is measured externally to high precision. This parameter is useful for studying a large range of physics, including dark energy~\citep{Linder:2005in}, modified gravity~\citep{Linder:2007hg} and the effects of neutrino mass~\citep{1983ApJ...274..443B}. Precision measurement of $f\sigma_8$ also allows using the kSZ to learn about astrophysics, such as the characteristics of the electron density profiles around halos, by breaking the degeneracy between the electron-scattering optical depth and the growth rate~\citep{Smith:2018bpn}. 

This paper is organised as follows: We begin with a description of kSZ and moving-lens tomography in Section~\ref{sec:vel_rec}. We show the dependence of the velocity field on cosmology and discuss the anticipated RSD bias in Section~\ref{sec:radial_velo_cosmo}. We introduce a formalism for the three-velocity in Section~\ref{sec:three_velocity_signal}. We introduce a Fisher analysis in Section~\ref{sec:fisher_ana}. We provide results from our forecasts using experimental specifications anticipated for next generation CMB and galaxy surveys in Section~\ref{sec:results}. We conclude with a discussion in Section~\ref{sec:discussion}. Throughout, we use the standard $\Lambda$CDM model with cosmological parameters $\{\Omega_b\, h^2,\,\Omega_{\rm cdm}h^2,\,A_s,\,n_s,\,\tau\}$ set equal to $\{0.022,0.12,{2.2\!\times\!10^{-9}},3.04,0.965,0.06\}$, respectively.

\section{Velocity reconstruction}\label{sec:vel_rec}

\subsection{kSZ tomography}\label{sec:ksztomography}

\begin{figure}[t!]
    \includegraphics[width=1\columnwidth]{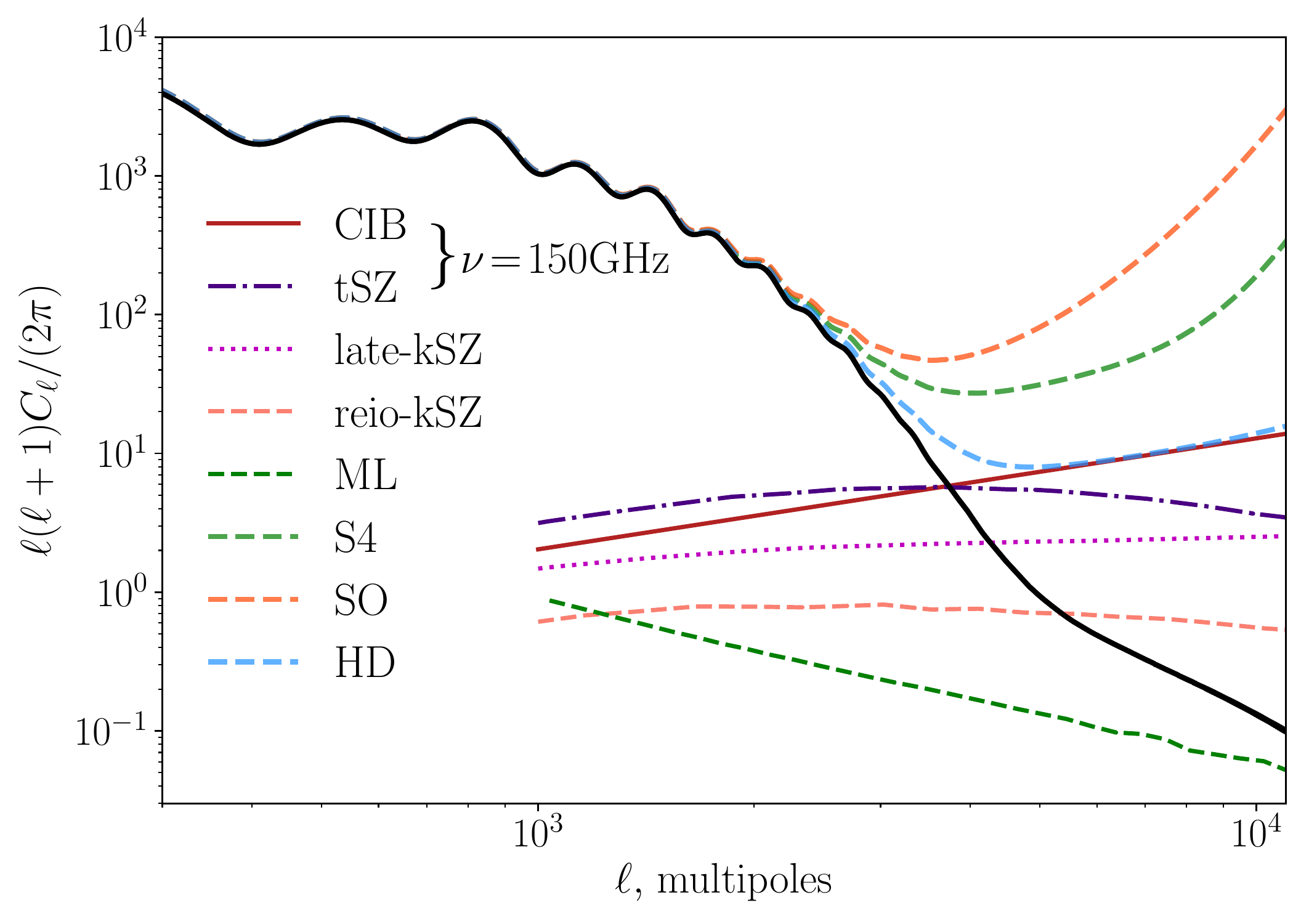}
    \vspace{-0.8cm}
    \caption{\textbf{CMB signal and foregrounds at the frequency $\boldsymbol{\nu=150}$ GHz}. Total (lensed) CMB signal is shown with the black solid line. We show the anticipated post-component separation noise (including foreground residuals) for three experiments; CMB-S4 (S4), Simons Observatory (SO) and CMB-HD (HD); along with the most significant foreground components (before foreground cleaning) in temperature maps (shown at 150 GHz). These are the cosmic infrared background (CIB) the thermal Sunyaev Zel'dovich effect (tSZ), late-time kSZ effect (due to free energetic electrons in intergalactic mediums where $z\in[0,6]$), the kSZ signal from patchy-reionization and the moving-lens (ML) effect. }\label{fig:signal_noiseCMB}
    \vspace{-0.3cm}
\end{figure}

Kinetic Sunyaev Zel'dovich tomography aims to extract cosmological information from the kSZ effect by using measurements of the CMB and a tracer of the electron density, such as a galaxy survey, to reconstruct the radial-velocity field. The temperature anisotropy induced by the kSZ effect from a large-scale structure (LSS) shell at redshift $z=z_*$ is
\be
\Theta_{\rm kSZ}(\boldsymbol{\theta})=K(z_*)\int_0^L\dd r\, q_\p(\boldsymbol{x}), 
\ee
where $\boldsymbol{x}\equiv\chi_\star\boldsymbol{\theta}+r\hat{\boldsymbol{r}}$, $\chi_\star$ is the conformal distance to the LSS shell, $\rhat$ is the unit vector in the radial direction, $\boldsymbol{\theta}$ is the angular direction on the sky, $\Theta(\boldsymbol{\theta})=\Delta T(\boldsymbol{\theta})/\bar{T}$ is the fractional CMB temperature fluctuation, and
\be
K(z)=-\sigma_T n_{e,0}x_e(z)e^{-\tau(z)}(1+z)^2\,,
\ee 
is the radial weight function with units of $1/$Mpc. Here, $\tau(z)$ is the optical depth at redshift $z$, $\sigma_T$ is the Thomson cross-section, $n_e$ is the free electron number-density and $q_\p(\boldsymbol{x})=\delta_e(\boldsymbol{x})v_\p(\boldsymbol{x})$ is the electron-momentum field, projected onto the radial direction. Most of the cosmological information is contained in the radial-velocity field $v_\p$ while $n_e$ depends primarily on astrophysics and non-linear large-scale structure; see Ref.~\cite{Smith:2018bpn} for a detailed discussion of this point. We demonstrate the anticipated late-time kSZ signal in Fig.~\ref{fig:signal_noiseCMB}, along with other foregrounds and the forecasted noise levels for next-generation CMB experiments. The component separated CMB noise due to foreground residuals, including the cosmic infrared background (CIB), radio sources, the thermal Sunyaev Zel'dovich effect (tSZ), and the kSZ signal are shown in Fig.~\ref{fig:signal_noiseCMB} for the Simons Observatory, CMB-S4 and CMB-HD experimental specifications (these were produced using the  \texttt{orphics}\footnote{\hyperlink{https://github.com/msyriac/orphics}{github.com/msyriac/orphics}} code as described in Appendix \ref{app:ilc}). We use the electron model described in Ref.~\citep{Battaglia:2016xbi}. 

The (inverse) noise on the velocity reconstruction is given by~\citep{Smith:2018bpn} 
\be\label{eq:ksz_tomo}
\frac{1}{N_\parallel(\boldsymbol{k}_L)}=\frac{K_*^2}{\chi_*^2}\int\!\!\frac{k_s\dd k_s}{2\pi}\!\left(\!\frac{P_{ge}(k_s)^2}{P^{\rm obs}_{gg}(k_s)C_\ell^{TT,\rm obs}}\!\right)_{\!\!\!\ell=k\chi_*}\!\!\!\!\,,\!\!\,\,\,\,\,\,
\ee
where $\bk$ is the three-dimensional Fourier wavevector and the integral is over small-scale Fourier modes $k_S$. We represent large-scale modes with an `$L$' subscript.  Here, 
$P_{ge}(k)$ is the power-spectrum of the galaxy-electron correlation, $P^{\rm obs}_{gg}(k)$ is the observed galaxy power spectrum and $C_\ell^{TT,\rm obs}$ is the observed CMB spectrum. We show the reconstruction noise from kSZ tomography in Fig.~\ref{fig:shot_on_weyl}.

\subsection{Moving-lens tomography}\label{sec:moving_lens_tomo}

Gravitational potentials that evolve in time induce a temperature modulation on the CMB known as the integrated Sachs-Wolfe (ISW) effect, which has the form
\be
\Theta_{\rm ML}(\boldsymbol{\theta})=-2\int_0^L\dd r\, \dot{\Phi}(\boldsymbol{x})\,,
\ee
where $\Phi$ is the gravitational potential. The ISW effect can be sourced by peculiar velocities of potentials, leading to the moving-lens effect, which has the form
\be\label{eq:main_1}
\begin{split}
% \Theta(\nhat)\!=\!-\frac{2}{c^2}\!\int\!\frac{\dd\chi}{c}\boldsymbol{\nabla}\Phi(\chi\nhat)\cdot\boldsymbol{v}_{\perp}(\chi\nhat)\,,
\Theta_{\rm ML}(\boldsymbol{\theta}) =-2\int_0^L\dd r\, \boldsymbol{\nabla}_{\!\perp} \!{\Phi}(\boldsymbol{x})\cdot \boldsymbol{v}_\perp(\boldsymbol{x})\,,
\end{split}
\ee
where $\boldsymbol{v}_{\perp}(\boldsymbol{x})$ is the peculiar (comoving) transverse velocity. Our focus is on the large-scale velocity field, where we anticipate that the true velocity is linear and curl-free, such that we can define a transverse velocity potential $\Upsilon(\boldsymbol{x})$ as $\boldsymbol{v}_\perp(\boldsymbol{x}) = \boldsymbol{\nabla}_\perp{\vpot}(\boldsymbol{x})$. We demonstrate the anticipated moving lens (ML) signal in Fig.~\ref{fig:signal_noiseCMB}. The ML signal is calculated using the velocity-reconstruction pipeline described in Ref.~\citep{upcoming:cayuso} for the redshift range $z\in\{0.1,5\}$. 

\begin{figure*}[t!]
    \includegraphics[width=2.1\columnwidth]{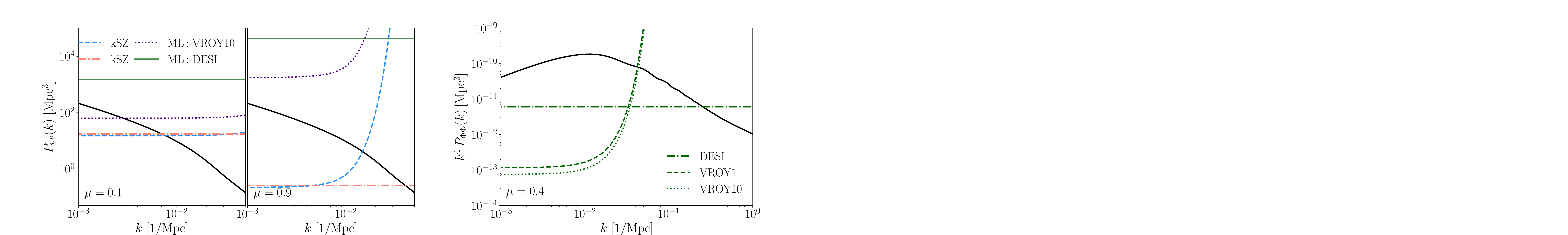}
    \vspace{-0.5cm}
    \caption{(\textbf{\textit{Left}})~ Comparison of the velocity signal to reconstruction noise at redshift $z=1$. The power spectrum of the norm  $|\boldsymbol{v}|$ of the 3-velocity is shown with solid black curves. Coloured lines correspond to reconstruction-noise spectra, calculated from the anticipated moving-lens (ML) and kSZ signals. Two galaxy-survey configurations are shown; these are DESI and Rubin Observatory Year 10 (VROY10). The CMB-S4 experimental specifications are used for all noise spectra. The two subplots correspond to nearly-transverse ($\mu=0.1$) and nearly-radial ($\mu=0.9$) velocity reconstruction. (\textbf{\textit{Right}})~The gravitational potential signal (solid black line) is shown together with the shot noise for the galaxy surveys we consider in this paper. In all plots, the rise in the noise on small-scales for Rubin is due to scatter from photometric redshifts.}\label{fig:shot_on_weyl}
    \vspace{-0.2cm}
\end{figure*}

Moving-lens tomography can be used to measure the large-scale transverse-velocity field. In the limit $\k\ll\k_S,\boldsymbol{\ell}/\chi_*$, we find the noise on the velocity reconstruction to be
\be
 \frac{1}{N^{\hat{\Upsilon}\hat{\Upsilon}}(\boldsymbol{k}_L)}\!&=&\!\frac{2\mu_\perp^2k_L^2}{\chi_*^2}\!\int\!\frac{k^3_S\,\dd k_S}{2\pi}\!\left(\!\frac{P_{\Phi\Phi}(k_S)^2}{P^{\rm obs}_{\Phi\Phi}(k_S)C_\ell^{\Theta\Theta,\rm obs}}\!\right)_{\!\!\!\ell=k_S\chi_*}\!\!\!\!\!\!,\non
\ee
where $\mu_\perp=\sqrt{1-\mu^2}$, $\mu=\boldsymbol{\hat{k}}\cdot\rhat$, $P_{\Phi\Phi}^{\rm obs}(k,\mu)=P_{\Phi\Phi}(k)+N_{\Phi\Phi}(k,\mu)$, and
\be
N_{\Phi\Phi}(k,\mu)={[(1+z)\rho_{m,0}/2]^2}/[{k^4\,b_g^2\,n_g\,W(k,\mu)^2}]\,,\,\,\,\,\,
\ee
with the photo-$z$ term, 
\be
W(k,\mu)=\exp\{-\mu^2k^2\sigma_z^2/2H_*^2\}\,,
\ee 
where $\sigma_z$ is the photo-$z$ error and $H_*$ is the Hubble parameter evaluated at redshift $z_*$. Here, $\rho_{m,0}=3\Omega_M H_0^2$, $\Omega_M$ is the matter fraction and $H_0$ is the Hubble constant. We show the shot noise on the gravitational potential in Fig.~\ref{fig:shot_on_weyl} for the galaxy surveys we consider, together with the reconstruction noise from moving-lens tomography. 

\section{Cosmology from velocities}\label{sec:radial_velo_cosmo}

\subsection{Cross-correlation signal} 

Where linear theory is valid, the reconstructed velocity fields are proportional to the cosmic growth rate. They satisfy
\be
\hat{v}_{\p}&=&b_\p\,\mu\frac{f a H}{k}\delta_m(\boldsymbol{k})\,,\\
\hat{v}_{\perp}&=&b_\perp\sqrt{1-\mu^2}\frac{f a H}{k}\delta_m(\boldsymbol{k})\,,
\ee  
where $H$ is the Hubble parameter, $a$ is the scale factor, $\delta_m(\boldsymbol{k})$ is the matter overdensity, $f=\dd\ln D(a)/\dd \ln a$ is the growth rate and the bias parameters $b_\p$ and $b_\perp$ are due to the mismatch between $P_{ge}^{\rm true}$ and $P_{ge}^{\rm fid}$, and similarly for $P_{gm}$. Here, $D(a)$ is the linear-theory growth factor for the matter spectrum that parametrises the time evolution of the matter power-spectra, $P_{mm}(a)=D^2(a)P_{mm}(a=1)$. The combination of the galaxy and the velocity satisfies
\be
P_{gg}(k,\mu)&=&(b_g+b_{\rm rsd}f\mu^2)^2P_{mm}(k)\label{eq:galaxy_spec}\,,\\
P_{gv}(k,\mu)&=&b\left(\frac{faH}{k}\right)(b_g+b_{\rm rsd}f\mu^2)P_{mm}(k)\,,\\
P_{vv}(k,\mu)&=&b^2\left(\frac{faH}{k}\right)^2P_{mm}(k)\,,
\ee
where $b\in\{b_\p\mu,b_\perp\!\sqrt{1-\mu^2}\}$ for the radial and the transverse velocities, respectively. Here, $b_g$ is the linear galaxy bias, $P_{gv}$ is the galaxy-velocity cross power spectrum and $P_{gg}$ is the galaxy auto
power spectrum. Above, we included the RSD effect~\citep{Kaiser:1987qv} $f\mu^2$ with a bias factor $b_{\rm rsd}$, which we discuss next. These equations demonstrate the dependence of observables on cosmological parameters.

\subsection{Redshift space distortions}\label{sec:fisher_ana}%\label{sec:rsd_bias}

In the absence of measurement biases (and on large-scales) the effect of RSDs is given by the Kaiser formula, which states that the galaxy profile $u_g$ is modified as
\be
u_g(\k)\rightarrow\left(1+\beta \mu^2\right)u_g(k)\,,
\ee
where $\beta=f/b_g$. Equivalently, the galaxy-density field in Fourier space takes the form 
\be
\delta_g(\k)=(b_g+\mu^2f)\delta_m(\k)+{\rm noise}\,. 
\ee
Extracting cosmological information from $f\delta_m(\k)$ assumes that one can calibrate a Fourier mode measured from the galaxy density field to the comoving $\k$, subject only to the knowledge of the distance scales, $D_a(z)$ and $H(z)$, for $k_\p$ and $k_\perp$, respectively. This can be done by using the BAO signature and the Alcock-Paczynski~\citep{Alcock:1979mp} method, internally for a galaxy survey, for example. In reality, however, the galaxies are not randomly distributed and their orientations show correlations with their local large-scale structure environment (tidal fields, density, dust, etc.). 

The appearance of the galaxy depends on the line-of-sight, expressed in the galaxy's frame of reference. This introduces an anisotropic selection function when measuring galaxy clustering, which depends on the intrinsic alignment of galaxies with its local environment~\citep{0903.4929}. In this work, we represent such effects with an RSD bias as
\be
\delta_g(\k)=(b_g+b_{\rm rsd}\mu^2f)\delta_m(\k)+{\rm noise}\,. \ee
Evidence for an RSD bias has recently been found at $5\sigma$~in Ref.~\citep{Obuljen:2020ypy} where authors find $b_{\rm rsd}$ variations on the order of $10-20\%$ within the BOSS CMASS and LOWZ samples~\citep{2017ApJS..233...25A}. Next, we demonstrate the improvement on cosmological constraints from the moving-lens effect in the presence of an RSD bias.

\section{Fisher analysis}\label{sec:three_velocity_signal}

Following~Ref.~\citep{Smith:2018bpn}, we use a simplified `snapshot' geometry throughout this paper to represent the observed Universe, which we take to be a 3D box with comoving side length $L$ and volume $V=L^3$, at a constant redshift $z_*=1$ corresponding to a distance $\chi_*\simeq3400$~Mpc. 

The optical-depth degeneracy can be represented on large scales with a bias on the radial-velocity measurement, in the form,
\be
\hv_\p(\k)= &&\,  b_\p v_\p^{\rm true}(\k)\\ 
&& + \big[ \mbox{noise with power spectrum $N_\p(\bk)$}\,
\big]\,,\nonumber 
\label{eq:hv_r}
\ee
where ${v}^{\rm true}_\p$ is the true radial-velocity amplitude. The radial velocity bias $b_\p$ depends only on the small-scale modelling of the electron-galaxy correlation. It can be calculated as~\citep{Smith:2018bpn}:
\be\label{eq:define_bias}
b_\p=\frac{\int\dd k_S F(k_S)P_{ge}^{\rm true}(k_S)}{\int\dd k_S F(k_S)P_{ge}^{\rm fid}(k_S)}\,,
\ee
where
\be
F(k_S)=k_S\frac{P_{ge}^{\rm fid}(k_S)}{P_{gg}^{\rm obs}(k_S)}\frac{1}{C_{\ell=k_S\chi}^{TT,\rm obs}}\,,
\ee
which depends on the difference between the true galaxy-electron correlation, $P_{ge}^{\rm true}$, and the fiducial model, $P_{ge}^{\rm fid}$. 

A similar bias can also appear in the transverse-velocity reconstruction, as:
\be
\hv_\perp(\k)=\,&& b_\perp v_\perp^{\rm true}(\k) \\
&& + 
\big[ \mbox{noise, with power spectrum $N_\perp(k,\mu)$} \big]\,,\nonumber \label{eq:hv_perp}
\ee
where ${v}^{\rm true}_\perp$ is the true transverse-velocity amplitude. For the moving-lens reconstruction, this bias is due to the uncertain cross-correlation between the gravitational potential and the density tracer. For a galaxy field tracer, for example, this bias can be written as: 
\be\label{eq:b_t}
b_\perp=\frac{\int\dd k_S F(k_S)P_{\Phi g}^{\rm true}(k_S)}{\int\dd k_S F(k_S)P_{\Phi g}^{\rm fid}(k_S)}\,,
\ee where $F(k_S)=k_S{P_{\Phi g}^{\rm fid}(k_S)}/[{P_{gg}^{\rm obs}(k_S)}{C_{\ell=k_S\chi}^{TT,\rm obs}}]\,.$ Here, $P_{\Psi g}^{\rm true}(k)$ is the true galaxy-potential correlation and $P_{\Psi g}^{\rm fid}(k)$ is the fiducial model.

We take the true velocity field to be curl-free, i.e.
\begin{equation}
v_j^{\rm true}(\k) = i \hk_j s(\k)  \label{eq:v_s}\,,
\end{equation}
where $s(\k)$ is a scalar field with power spectrum 
$P_{vv}(k) = (faH/k)^2 P_{mm}(k)$. To avoid covariance matrices which are not full rank,
we combine $\hat v_\p$ and $\hat v_\perp$
into a ``unified'' reconstruction $\hat v_i$
(a three-vector).
Then, the two-point function of $\hat v_i$ can be written as the
sum of signal and noise contributions from the velocity, as:
\begin{equation}\label{eq:signal_and_noise_v}
\big\langle \hat v_i(\k) \, \hat v_j(\k')^* \big\rangle 
= \Big( S_{ij}(\k) + N_{ij}(\k) \Big) \, (2\pi)^3 \delta^3(\k-\k')\,.
\end{equation}
The average (shown with brackets) is over realisations of the three-velocity.

Given a wavenumber $\k_i$ and line-of-sight direction $\hr_i$,
one natural choice of basis for the 3-by-3 matrices
$S_{ij}$ and $N_{ij}$ is the ``radial/tangential'' basis, where $\hr_i$ is a basis
vector:
\begin{equation}
\hr_i = \left( \begin{array}{c} 1 \\ 0 \\ 0 \end{array} \right)
 \hspace{1cm}
\k = \left( \begin{array}{c} k \mu \\ k \sqrt{1-\mu^2} \\ 0 \end{array} \right)\,.
 \hspace{1cm}  \label{eq:rt}
\end{equation} In what follows, we add the galaxy field to Eq.~\eqref{eq:signal_and_noise_v}, promoting $S_{ij}(\k)$
and $N_{ij}(\k)$ to 4-by-4 matrices.   

%\section{Fisher analysis}

In the radial/tangential basis, the noise covariance [including the galaxy field, $g(\boldsymbol{k})$] can be read off
from Eqs.~(\ref{eq:hv_r})~and~(\ref{eq:hv_perp}):
\be
&& N_{ij}(\k)\\
&& ={\rm diag}\{N_\p(k,\mu),N_\perp(k,\mu),N_\perp(k,\mu),1/[W^{2}(k,\mu)n_g]\}\nonumber\label{eq:Nij_RT}\,,
\ee
where $n_g$ is the galaxy number-count density in a given box. To compute the signal covariance, we combine 
Eqs.~(\ref{eq:hv_r})~and~(\ref{eq:hv_perp}) to get
\begin{equation}
\hv_i(\k) = B_{ij}(\k) v_j^{\rm true}(\k) + (\mbox{noise})\,,
\end{equation}
where the 4-by-4 bias matrix $B_{ij}(\k)$ is given by
\be
B_{ij}(\k) ={\rm diag}\{b_\p,b_\perp,b_\perp,b_g\}\,.
\ee
By Eq.~(\ref{eq:v_s}), the signal covariance of $v_i^{\rm true}$
is
\begin{equation}
\big\langle v_i^{\rm true}(\k) v_j^{\rm true}(\k')^* \big\rangle = 
 \hk_i \hk_j P_v(k) \, (2\pi)^3 \delta^3(\k-\k')\,,
\end{equation}
and the signal covariance of $\hv_i$ is
\be\label{eq:Sij_RT}
S_{ij}=c_i c_j P_{vv}(k)\,,
\ee
where
\be
c_i&=&(c_\p,c_\perp,0,c_g)\,,\\
c_\p&\equiv&b_\p\mu\,,\\
c_\perp&\equiv&b_\perp \sqrt{1-\mu^2}\,,\\
c_g&\equiv&(b_g+b_{\rm rsd} f \mu^2)\, {k}/({faH})\,.
\ee

Given parameters $\pi_1, \cdots, \pi_N$, the $N$-by-$N$
Fisher matrix is
\begin{equation}
F_{ab} = \frac{V}{2} \int \frac{d^3\k}{(2\pi)^3} \, 
  \mbox{Tr} \left[ 
      C^{-1}(\k) \frac{\partial S(\k)}{\partial\pi_a}
      C^{-1}(\k) \frac{\partial S(\k)}{\partial\pi_b}
  \right]\,,
\end{equation}
where $C=S+N$ is the total covariance.
We can simplify a little by noting that the 
signal and noise covariance matrices in
Eqs.~(\ref{eq:Nij_RT})~and~(\ref{eq:Sij_RT})
only depend on $\k$ through $(k,\mu)$.
This lets us reduce the 3D integral to a 2D integral,
\be\label{eq:Fisher}
F_{ab} = \frac{V}{2}&& \int_0^\infty  \frac{k^2\, dk}{4\pi^2}  
  \int_{-1}^1 d\mu \\
  \times&&\mbox{Tr} \left[ 
      C^{-1}(k,\mu) \frac{\partial S(k,\mu)}{\partial\pi_a}
      C^{-1}(k,\mu) \frac{\partial S(k,\mu)}{\partial\pi_b}
  \right]\,. \nonumber
\ee
In our forecasts we consider the following 6 parameters: $\{f\sigma_8,\,b_g\sigma_8,\,b_\p,\,b_\perp,\,b_{\rm rsd}\}$.

\section{Results}\label{sec:results}

\begin{figure}[t!]
    \includegraphics[width=\columnwidth]{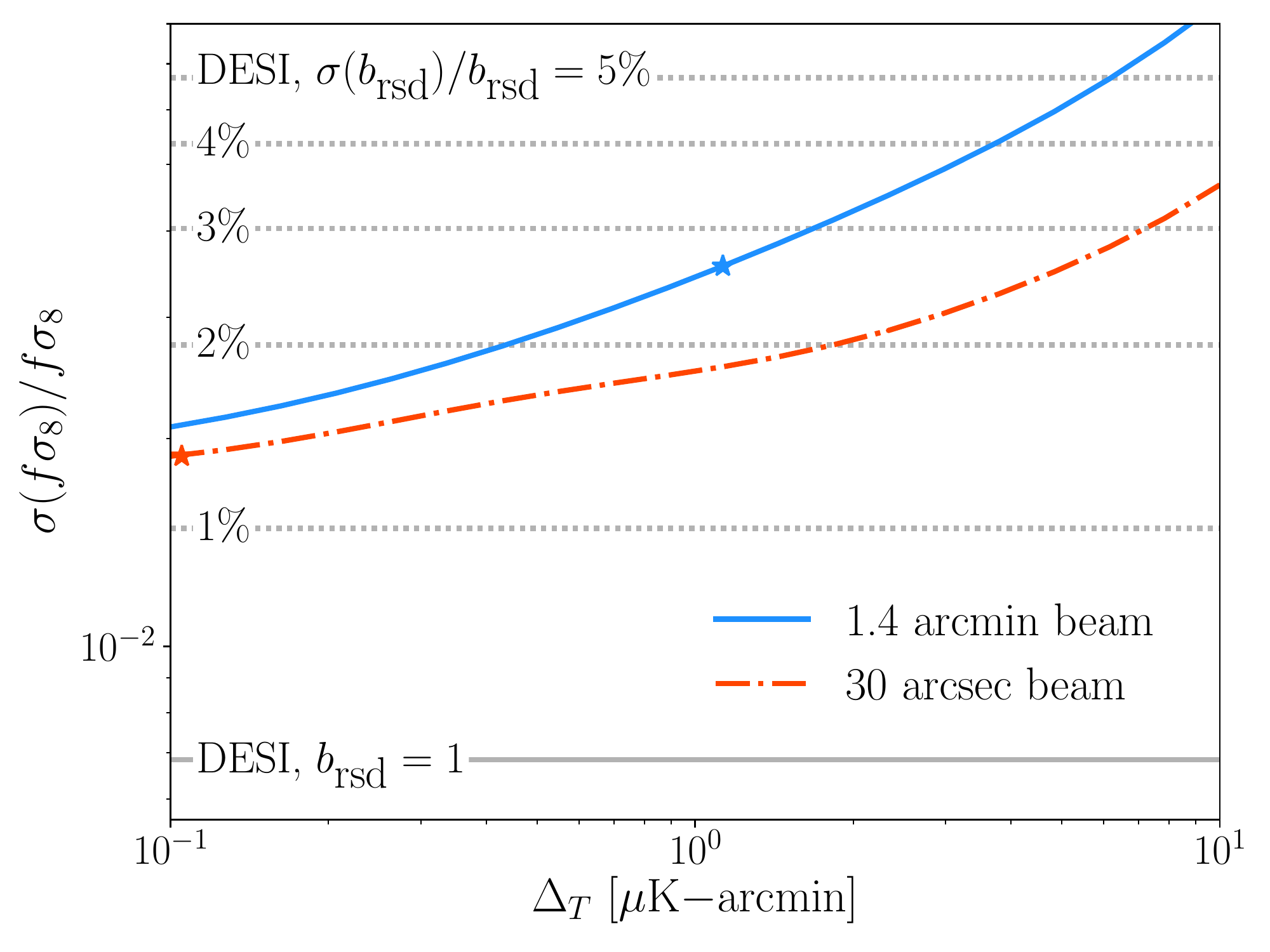}
    \vspace{-1cm}
    \caption{\textbf{Constraints on $\boldsymbol{f\sigma_8}$ from moving-lens tomography and galaxy clustering.} Coloured lines correspond to constraints from the reconstructed transverse-velocity fields, and their cross-correlation with the galaxy field on large scales (with no RSD bias prior). A range of CMB experiment noise levels ($\Delta_T$) are shown and two beam sizes, appropriate for CMB-S4 and CMB-HD. We scale the noise levels for all of the frequency bands of these experiments. The ILC calculation is discussed in Appendix~\ref{app:ilc}. Blue (orange) markers show anticipated noise levels of CMB-S4 (CMB-HD). We use Rubin Observatory Year 10 (VROY10) for moving-lens tomography. Note that here the radial-velocity reconstruction only marginally improves the constraints on $f\sigma_8$: the constraints are dominated by the correlation of the transverse-velocity field and galaxy clustering. We consider a $1\%$ prior on the transverse-velocity bias [defined in Eq.~\eqref{eq:b_t}], from an external measurement of the matter-galaxy correlation on small scales, such as galaxy and lensing correlations, for example. {The $1\%$ constraints on the transverse-velocity bias falls near the scenario where $b_\perp$ is fixed and not marginalised.} Finally, we show constraints on the $f\sigma_8$ parameter from measurements of the RSD effect, using the spectroscopic DESI galaxy survey. The dotted gray horizontal lines corresponds to $5\%-1\%$ priors on the RSD bias from top to bottom. The solid gray horizontal line correspond to fixing $b_{\rm rsd}$ to 1. Recent studies show evidence for the case $b_{\rm rsd}\neq1$~\citep{Obuljen:2020ypy}. }\label{fig:fs8_from_lssty10_only_tomography}
    \vspace{-0.2cm}
\end{figure}

\begin{figure}[b!]
    \includegraphics[width=\columnwidth]{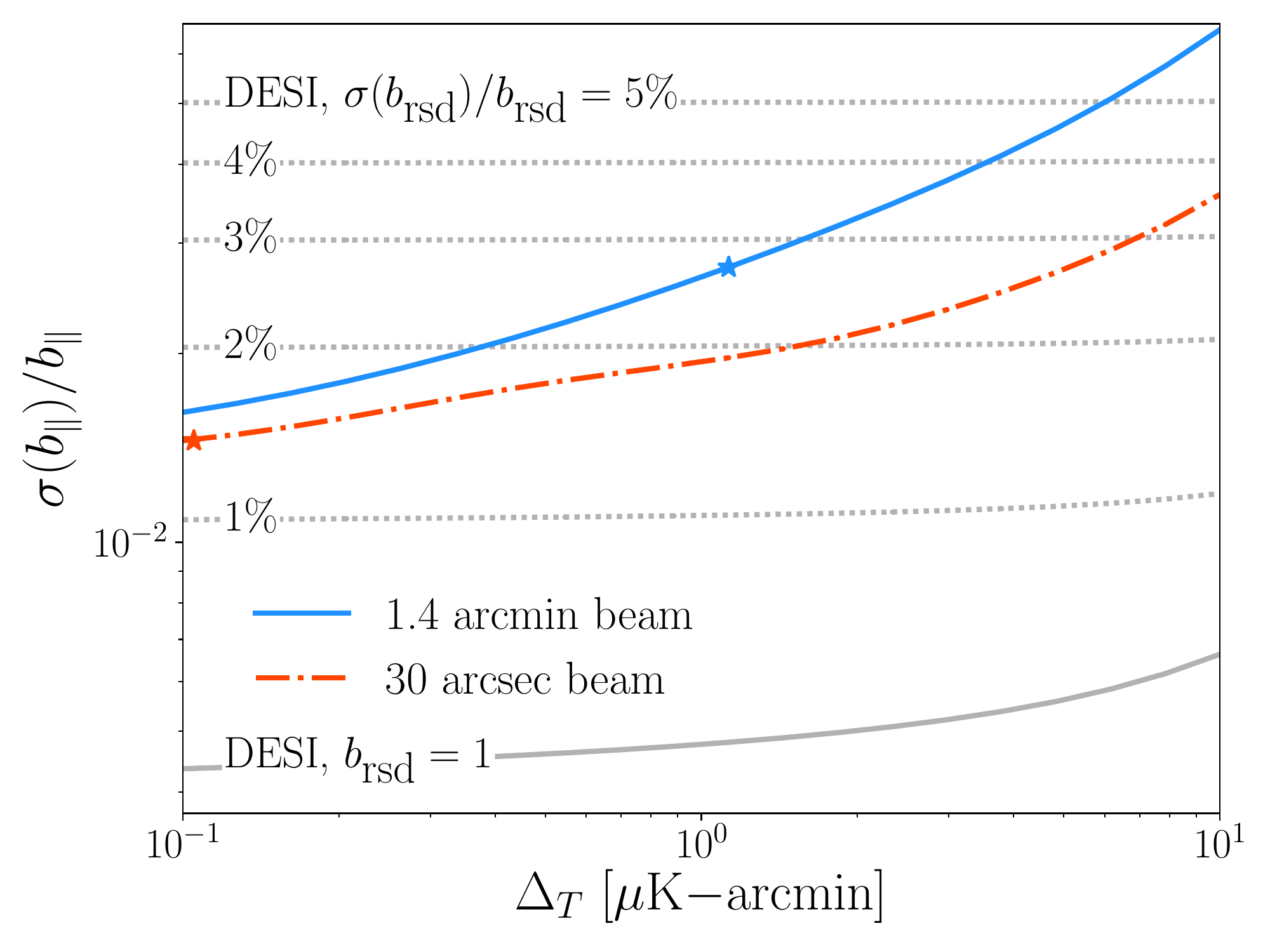}
    \vspace{-1cm}
    \caption{\textbf{Constraints on $\boldsymbol{b_\p}$ from combination of moving-lens tomography, kSZ tomography and galaxy clustering.} Coloured lines correspond to constraints from the reconstructed transverse- and radial-velocity fields, and their cross-correlation with the galaxy field on large scales (with no RSD bias prior); similar to Figure~\ref{fig:fs8_from_lssty10_only_tomography}, with  $1\%$ prior on the transverse-velocity bias. We use Rubin Observatory Year 10 (VROY10) survey for kSZ and moving-lens tomography. Gray lines correspond to constraints on $b_\p$ from measurements of the RSD effect and kSZ tomography, using the spectroscopic DESI galaxy survey and the CMB-S4 experimental specifications. The dotted gray lines corresponds to $5\%-1\%$ priors on the RSD bias from top to bottom. The solid grey line correspond to fixing $b_{\rm rsd}$ to 1.}\label{fig:moving_lens_brsd_alt_br}
    \vspace{-0.2cm}
\end{figure}

We consider a range of CMB experiments with varying noise, including SO and CMB-S4, as well as the futuristic CMB-HD survey. The galaxy-survey (shot) noise depends on the number-density function, $n_g(z)$. We consider three LSS experiments: DESI, Vera Rubin Observatory Year 1 (VROY1) and Year 10 (VROY10). The parameters for VRO are based on~\citep{2012JCAP...04..034H,LSSTDarkEnergyScience:2018jkl} and those for DESI are based on~\citep{DESI:2016fyo}. For DESI, we take a 116~Gpc$^3$ box with galaxy survey density $n_g=1.4\times10^{-4}/$Mpc$^3$ and $\sigma_z=0$. For VROY1, we take a 113.4~Gpc$^3$ box, with galaxy survey density $n_g=6.9\times10^{-3}/$Mpc$^3$. For VROY10, we take a 180Gpc$^3$ box, with galaxy survey density $n_g=1.2\times10^{-2}/$Mpc$^3$. For both VROY1 and VROY10, we consider photo-$z$ errors satisfying $\sigma_z=0.06$ at $z=1$. For $\{$DESI,\,VROY1,\,VROY10$\}$, we consider a galaxy bias of $\{1.5,1.7,1.6\}$, respectively. We take all the boxes to be centered at $z=1$. For all forecasts, we set the maximum wavenumber we include in our analysis to $k_{\rm max}=0.1/$Mpc.
 
The total CMB power spectrum gets contributions from weak gravitational lensing, the kSZ effect (both from reionization and late times) and the experimental noise, which we take to satisfy 
\be
N_\ell=\Delta_T^2\exp\left[\frac{\ell(\ell+1)\theta^2_{\rm FWHM}}{8\ln2}\right]\,.
\ee
Note that when forecasting the CMB noise for these experiments, we also include the contribution to the power spectra from residual foregrounds after foreground cleaning we discuss in Appendix~\ref{app:ilc}. Fig.~\ref{fig:signal_noiseCMB} demonstrates how the CMB noise and different residual foregrounds compare with the CMB signal.

\begin{table}[b!]
  \begin{center}
    \caption{The detection signal-to-noise (SNR) of the (reconstructed) velocity and galaxy-density cross-correlation: $P_{\hat{v}g}(k)$. Velocities are reconstructed from the kSZ and moving-lens (ML) tomography for various CMB and LSS experiments. VROY1 and VROY10 refer to Vera Rubin Observatory Year 1 and 10 respectively.}
    \label{tab:detection-SNR}
    \begin{tabular}{| c | c c c | c | c c c |} 
    \hline
    kSZ SNR &  &  CMB &  & ML SNR &  &  CMB & \\ 
    \hline
    LSS & ~SO & S4 & HD & LSS & ~SO & S4 & HD~\\
   \hline
   DESI & 231 & 414 & 1170 & DESI & 7 & 8 & 62  \\
   VROY1 & 116 &
   210 &
   669  & VROY1 & 16 & 28 & 85 \\
   VROY10 & 123 & 228 & 775 & VROY10 & 20 & 37 & 112 \\ 
   \hline
   \end{tabular}
  \end{center}
  \vspace{-0.1cm}
\end{table}

In Table~\ref{tab:detection-SNR}, we show the detection signal-to-noise ratio (SNR) of the reconstructed velocity and galaxy density cross-correlation $P_{\hat{v}g}(k)$ from our experimental configurations. The velocities are reconstructed either form the moving-lens (ML) effect or the kSZ tomography. We find similar SNR forecasts compared to previous results~\citep{Hotinli:2018yyc,Hotinli:2020ntd}. Nevertheless, note that our calculation of the transverse velocity noise differs from earlier work by the use of the simplified box formalism here.\footnote{Our forecasts for the moving-lens SNR differs from Refs.~\citep{Hotinli:2018yyc,Hotinli:2020ntd}, in various ways: First, Refs~\citep{Hotinli:2018yyc,Hotinli:2020ntd} included a map-based reconstruction on the two-sphere and with 11 redshift bins in the range $z\in[0,3]$, unlike our single-box formalism which is simpler but potentially more optimistic than the former. Furthermore, forecasts in Refs~\citep{Hotinli:2018yyc,Hotinli:2020ntd} did not include the adverse effects of residual foregrounds after foreground cleaning (except the effects of kSZ and weak lensing), while including \textit{delensed} CMB spectra with ideal noise. Delensing noticeably improves the detection prospects of reconstruction at the low-noise limit, hence important for experiments like CMB-S4 and CMB-HD, for example. Despite these differences, we find matching SNR forecasts for CMB-S4 and SO.}

We show forecasts on $f\sigma_8$ in Fig.~\ref{fig:fs8_from_lssty10_only_tomography}. We find that moving-lens tomography, together with the galaxy density, can constrain $f\sigma_8$ to high precision, comparable to the scenario where the RSD bias is known up to few percent accuracy. For SO and VROY10, improvements on $f\sigma_8$ relative to RSD can be achieved if the RSD bias is uncertain at the level of $\sigma(b_{\rm rsd})>5\%$. Using S4 and VROY10 further improves the benefit from using the transverse velocities, while constraints from CMB-HD and VROY10 improves upon the scenario where the RSD bias is known up to around $1-2\%$ accuracy. 

Note that in our forecasts we considered a $1\%$ prior on the transverse velocity field. This can potentially be provided from external measurements of the galaxy-matter cross-correlation on small scales (from cross-correlation of galaxy-lensing measurements, for example) as defined in Eq.~\eqref{eq:b_t}. For upcoming experiments, we find $\lesssim1\%$ prior on the transverse-velocity bias recovers the cosmic-variance limit with the transverse-velocity bias fixed as $b_\perp=1$. 

Fig.~\ref{fig:moving_lens_brsd_alt_br} shows the constraints on $b_\p$ from a combination of moving-lens tomography, kSZ tomography, and galaxy clustering. Moving-lens measurement of $f\sigma_8$ removes the degeneracy between the radial-velocity bias $b_\p$ and $f\sigma_8$, suffered by the radial velocities reconstructed with kSZ tomography. This allows kSZ to measure directly $b_\p$, allowing the possibility of astrophysical inference. 

\begin{figure}[t!]
    \includegraphics[width=1\columnwidth]{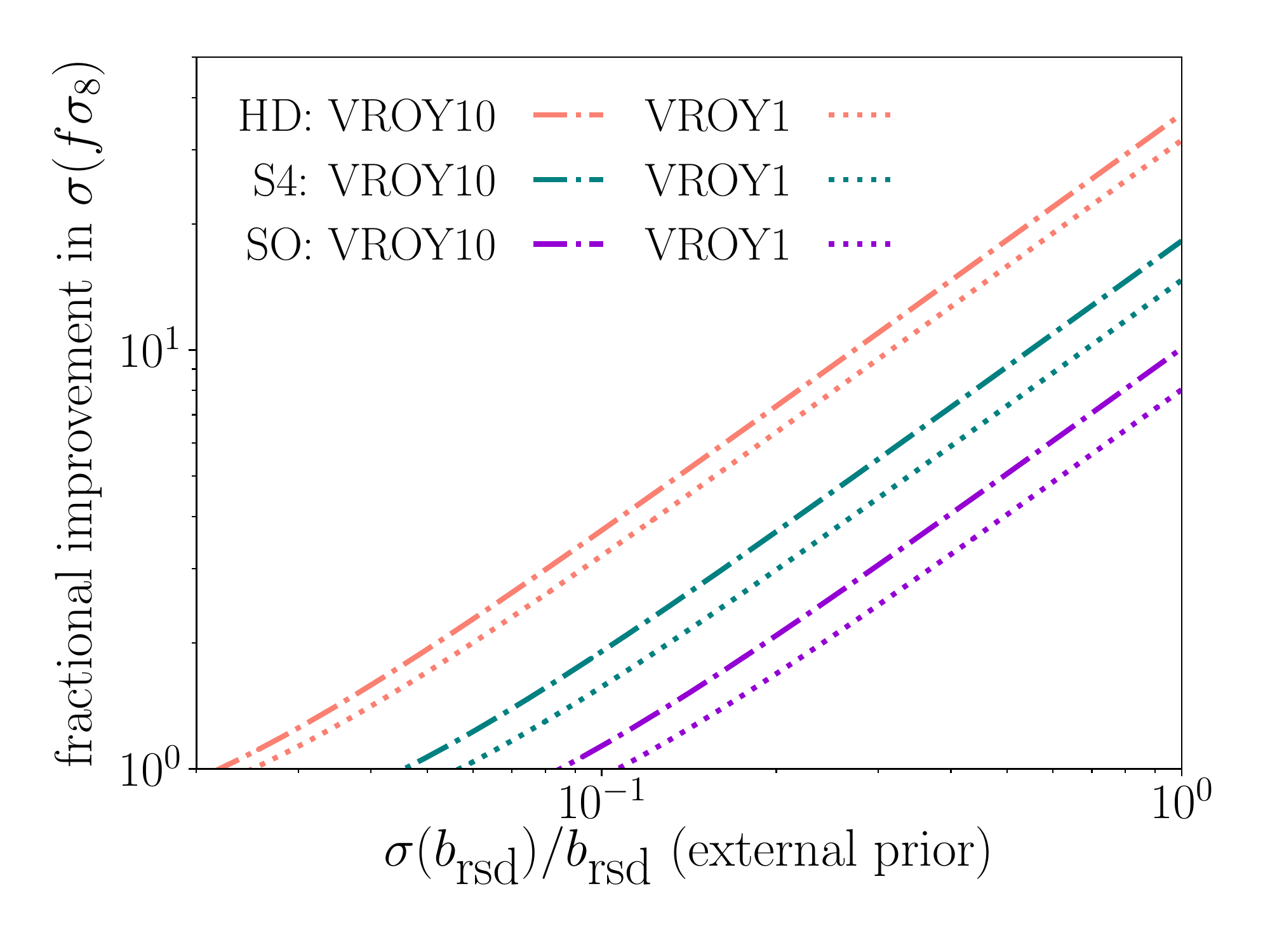}
    \vspace{-1cm}
    \caption{\textbf{The fractional improvement in $\boldsymbol{\sigma(f\sigma_8)}$ from the moving-lens effect, relative to the constraint from biased galaxy RSDs.} The ratio of constraints from DESI RSDs (with varying RSD bias prior) and the combined measurement of biased galaxy RSDs and transverse velocities is shown for various experimental configurations.} \label{fig:benefit_from_tomo_cmb_rsd}
    \vspace{-0.3cm}
\end{figure}

In Fig.~\ref{fig:benefit_from_tomo_cmb_rsd}, we show the ratio of the errors on $f\sigma_8$ obtained from the DESI RSDs and from forecasts combining VRO galaxies and the moving-lens tomography. The RSDs are taken to be biased for both calculations with a varying prior assumption on $\sigma(b_{\rm rsd})$ (shown on the $x$~axis). Our results suggest that moving-lens tomography improves the $f\sigma_8$ constraints for $\sigma(b_{\rm rsd})/b_{\rm rsd}\gtrsim{\rm few}\%$. Note that in the absence of prior knowledge on the RSD bias [or for $\sigma(b_{\rm rsd})\sim5\%$ or worse], moving-lens tomography may play an essential role in our ability to constrain cosmology from velocity reconstruction. 
So far, the only other observable that can constrain $f\sigma_8$ (together with the kSZ tomography) is the FRB dispersion measurements, which allow the breaking of the optical-depth degeneracy of the kSZ signal~\citep{Madhavacheril:2019buy}. Our forecasts suggest  the moving-lens effect may provide a constraining power comparable to measurement of $\sim\!10^5$ FRBs, depending on the uncertainty on the FRB dispersion measurement~\citep{Madhavacheril:2019buy}.

\section{Discussion}\label{sec:discussion}

In this work, we have explored one of the cosmological applications of reconstructing the transverse-velocity field from the moving-lens effect. Velocity fields are direct tracers of the combination of cosmic growth rate and the amplitude of matter fluctuations and contain valuable cosmological information. While velocity measurements from the kSZ and the moving-lens effects are biased due to uncertain electron-galaxy and matter-galaxy cross-correlation on small-scales, respectively, the latter degeneracy can be broken by relatively robust measurements such as correlations of the galaxy density and (galaxy or CMB) weak lensing. Overcoming the former ``optical depth'' degeneracy, however, is likely more difficult. Galaxy clustering is also sensitive to cosmic growth due to the RSD effect~\citep{Kaiser:1987qv}, but future measurements may be biased by selection effects. Using moving-lens reconstructions that combine the Vera Rubin Observatory and CMB surveys  on the other hand provides strong constraints on cosmological parameters such as $f\sigma_8$ at the level of $\lesssim5\%$, $3\%$ and $2\%$ for the Simons Observatory, CMB-S4 and CMB-HD experiments respectively.

We focused on systematics in the RSD effect, but moving forward, a more detailed analysis involving the modelling of all relevant effects will give further insight into the benefit of moving-lens tomography. These effects can be separated into two categories. First, the theory predictions from our models are often gauge dependent and defined in comoving coordinates, rather than in redshift space. Mapping from the theory to gauge-independent observables (such as the galaxy power spectrum or CMB temperature) requires a multitude of correction terms on top of the density fluctuations, which take into account the photon geodesics in an inhomogeneous Universe. These include general relativistic effects such as Doppler (magnification) terms, potential (Sachs-Wolfe, integrated Sachs-Wolfe, time delay) terms and weak lensing, as well as anisotropies in the mapping from real space to redshift space induced by peculiar velocities (RSDs). Second, galaxy clustering measurements are subject to selection effects, where detecting particular galaxies with various properties may be more or less likely than on average. These effects include tidal alignments~\citep{0903.4929}, lensing magnification~\citep{Hui:2007cu}, and redshift-evolution biases~\citep[e.g.][]{Alonso:2015uua}. These biases can complicate the mapping between theory and observed quantities and should be explored in future work.

The upcoming decade will host a wealth of new and high-quality data. These developments will open many novel opportunities for cosmological inference. In particular, using the CMB as a cosmological backlight, secondary fluctuations induced by the interaction between CMB photons and large-scale structure allow new methods like transverse-velocity reconstruction from the moving-lens effect, as described in this paper. These new observables provide the different tracers that can potentially improve the constraints on cosmological parameters, beyond the limits imposed by selection effects, astrophysical uncertainties, and cosmic variance.

\section{Acknowledgements}

We thank J. Colin Hill for comments and suggestions on the ILC noise calculation. SCH is supported by the Horizon Fellowship from Johns Hopkins University. SCH was also supported by a postdoctoral grant from Imperial College London and a Perimeter Visiting Fellowship. SCH thanks Matthew Johnson and Juan Cayuso for useful conversations. MK was supported by NSF Grant No.\ 1818899 and the Simons Foundation. Research at Perimeter Institute is supported in part by the Government of Canada through the Department of Innovation, Science and Industry Canada and by the Province of Ontario through the Ministry of Colleges and Universities.

\bibliography{main}

\appendix
\section{Internal Linear Combination noise curves}
\label{app:ilc}

\begin{table}[h]
\begin{tabular}{|l|l|l|l|l|l|l|}
\hline
        & \multicolumn{3}{c|}{Beam FWHM}                                              & \multicolumn{3}{c|}{Noise RMS}                                              \\
        & \multicolumn{3}{c|}{}                                                       & \multicolumn{3}{c|}{($\mu$K-arcmin)}                                        \\ \cline{2-7} 
        & \multicolumn{1}{c|}{SO} & \multicolumn{1}{c|}{S4} & \multicolumn{1}{c|}{HD} & \multicolumn{1}{c|}{SO} & \multicolumn{1}{c|}{S4} & \multicolumn{1}{c|}{HD} \\ \hline
39 GHz  & $5.1'$                  & $5.1'$                  & $36.3''$                & 36                      & 12.4                     & 3.4                     \\
93 GHz  & $2.2'$                  & $2.2'$                  & $15.3''$                & 8                       & 2.0                     & 0.6                     \\
145 GHz & $1.4'$                  & $1.4'$                  & $10.0''$                & 10                      & 2.0                     & 0.6                     \\
225 GHz & $1.0'$                  & $1.0'$                  & $6.6''$                 & 22                      & 6.9                     & 1.9                     \\
280 GHz & $0.9'$                  & $0.9'$                  & $5.4''$                 & 54                      & 16.7                    & 4.6                     \\ \hline
\end{tabular}
\caption{{\it Inputs to ILC noise: } The beam and noise RMS parameters we assume for survey configurations roughly corresponding to Simons Observatory (baseline), CMB-S4 and CMB-HD. We do not include the 30 GHz channel and do not include atmospheric noise since the kSZ and moving lens information is primarily in scales $\ell>2000$.}
\label{tab:beamnoise}
\end{table}

We describe here the procedure used to analytically calculate the effective total noise in CMB maps for the Simons Observatory (SO), CMB-S4 and CMB-HD surveys, guided by the Internal Linear Combination (ILC) algorithm that would be used in a realistic analysis. Our approach here is approximate and the resulting noise curves are only meant to be roughly representative of the forecast performance of these surveys; we do however find 10-20\% agreement of the SO and CMB-S4 results with the official noise curves in \cite{Ade:2018sbj} and \cite{1907.04473}, respectively. In particular, for SO, we find better than 10\% agreement with the official forecasts for almost all relevant multipoles. We employ this analytic approach (as opposed to the simulation-based work in \cite{Ade:2018sbj} and \cite{1907.04473}) so as to be able to easily propagate the results of varying the white noise RMS in the CMB experiments.

We assume that each of these surveys measures the millimeter sky at 39, 93, 145, 225  and 280 GHz with beam FWHM and white noise RMS as shown in Table \ref{tab:beamnoise}. We do not include large-scale atmospheric noise but do not expect this to make a difference for our forecasts given the availability of {\it Planck} data at 44, 100, 143 and 217 GHz as well as the fact that most of the kSZ and moving lens information is derived from small scales with $\ell>2000$. 

For the signal contributions in the millimeter sky, we include contributions from tSZ, clustered CIB and conservative (large) levels of Poisson CIB foregrounds at these frequencies as well as the black-body late-time kSZ following the approach in \cite{1708.07502}, which is based on fits to ACT data from \cite{1301.0776}, but we do not include the tSZ-CIB correlation. We include radio sources in the 39, 93 and 145 GHz channels using the flux-limit-dependent radio source power model from \cite{Lagache19}, where for both SO and S4 we assume flux limits in those channels of 10, 7 and 10 mJy respectively. For HD, we assume lower flux limits of 2, 1 and 1 mJy respectively. In addition, we include the black-body lensed CMB contribution calculated using CAMB \cite{CAMB} as well as the reionization kSZ signal from \cite{1301.3607}. We do not include the contribution from the moving lens signal itself since it is sub-dominant to the sum of the above.

Given the above sky model involving signal components correlated across frequencies indexed by $i$ (and uncorrelated beam-deconvolved noise), the total covariance between two frequency channels at each multipole $\ell$ is ${\boldsymbol C}^{ij}_{\ell}$. The final minimum-variance (standard ILC) noise for the black-body CMB+kSZ signal that includes all contributions (including the sample variance in the CMB and kSZ) is then given by $N_{\ell} = \left[\sum_{ij}  \left({\boldsymbol C}^{-1}\right)^{ij}_{\ell}\right]^{-1}$. These curves are shown in Figure \ref{fig:signal_noiseCMB}.
\end{document}